\documentclass[intlimits,twoside,a4paper]{article}

\usepackage{amsmath,amssymb}
\usepackage{graphicx}
\usepackage{color}

\usepackage[T2A]{fontenc}
\usepackage[cp1251]{inputenc}

\usepackage[eqsecnum]{cmpj2}

\issue{2017}{20}{1}{13003}
\doinumber{10.5488/CMP.20.13003}

\title[Geometric characteristics of quantum evolution]%
{Geometric characteristics of quantum evolution: curvature and torsion %
}
\author[H.P. Laba, V.M. Tkachuk]{H.P. Laba\refaddr{label1}, V.M. Tkachuk\refaddr{label2}}
\addresses{
\addr{label1} Department of Applied Physics and Nanomaterials Science, Lviv Polytechnic National University,\\ 5 Ustiyanovych St., 79013 Lviv, Ukraine
\addr{label2} Department for Theoretical Physics,
Ivan Franko National University of Lviv,\\
12 Drahomanov St., 79005 Lviv, Ukraine
}

\authorcopyright{H.P. Laba, V.M. Tkachuk, 2017}
\date{Received January 20, 2017, in final form March 2, 2017}

\begin{document}

\maketitle

\begin{abstract}
We study characteristics of quantum evolution which can be called
curvature and torsion. The curvature shows a deviation of the
state vector in quantum evolution from the geodesic line.
The torsion shows a deviation of state vector from the plane of
evolution (a two-dimensional subspace) at a given time.
\keywords curvature, torsion, quantum evolution,
geometry of quantum state space
\pacs 03.65.-w, 03.65.Aa
\end{abstract}

\section{Introduction}
Geometric ideas play an important role in quantum mechanics, in
particular in the studies of  quantum evolution
\cite{Anandan90,Abe93,Grigorenko91,Kuzmak16}, quantum brachistochrone
problem \cite{Mosta07,Bender08,Fryd08,Kuzmak15}, entanglement of quantum states
\cite{Brody01,Brody07}, quantum correlations \cite{Deb16}, Berry phase that has geometric origin \cite{Shapere89}.

In the classical case, the curvature and
torsion are important geometric characteristics of the trajectory.
The aim of the present paper is to answer the question: What is
the quantum analogue of these classical geometrical notions?
Partly, the answer to this question was given in \cite{Brody96} where the authors from the perspective different from this paper, namely, considering geometry of quantum statistical interference, derived an explicit expression for the curvature of quantum evolution.

First, let us consider some facts of the geometry of the space of
quantum states. A distance between two quantum states
$|\psi_1\rangle$, $|\psi_2\rangle$ which is normed to unity can be defined in different
ways. In this paper, we will refer to the Fubiny-Study distance and
the Wootters distance defined, respectively, as follows:
\begin{eqnarray}
d^{({\rm
FS})}(|\psi_1\rangle,|\psi_2\rangle)=\gamma\sqrt{1-|\langle\psi_1|\psi_2\rangle|^2},
\\
d^{({\rm W
})}(|\psi_1\rangle,|\psi_2\rangle)=\gamma\arccos|\langle\psi_1|\psi_2\rangle|,
\end{eqnarray}
where $\gamma$ is an arbitrary constant (for a
short review see, for instance, \cite{Dodonov 99}).

These distances are equivalent for the neighboring states when
$|\langle\psi_1|\psi_2\rangle|^2=1-\delta^2$, where $\delta$ is a
small value, namely, $d^{({\rm FS})}=d^{({\rm W})}=\gamma\delta$.
 As a result, the element of length for the family (set) of quantum state vectors
$|\psi(\xi^1,\xi^2,\ldots,\xi^k)\rangle$ parameterized by $k$
parameters $\xi^1$, $\xi^2$, $\ldots\,\,$, $\xi^k$ is the same for different
definitions of the distance
\begin{eqnarray}
\rd s^2=g_{ij}\rd\xi^i\rd\xi^j
\end{eqnarray}
with metric tensor
\begin{eqnarray} \label{metric}
g_{ij}=\gamma^2 \Re(\langle\psi_i|\psi_j\rangle-\langle\psi_i|\psi\rangle\langle\psi|\psi_j\rangle),
\end{eqnarray}
where
\begin{eqnarray}
|\psi_i\rangle={\partial\over\partial\xi^i}|\psi(\xi^1,\xi^2,\ldots,\xi^k)\rangle.
\end{eqnarray}
This form of metrics of quantum states was discussed by many
authors (see, for example, \cite{Abe93,Provost80,Revicule97,Aalok06,Aalok07,Tka11}).

It is convenient to put $\gamma=2$. Then, in a two-dimensional case,
$g_{ij}$ is a metric tensor of a sphere with the radius equal to one
(the Bloch sphere).

According to
Schr\"odinger equation, one can introduce the velocity of quantum
evolution \cite{Anandan90}
\begin{eqnarray} \label{velocity}
v={\rd s\over \rd t}={\gamma\over\hbar}{\sqrt{\langle(\Delta H)^2\rangle}},
\end{eqnarray}
where $\Delta H=H-\langle H\rangle$.

\section{Geodesic in the space of quantum state vectors}
The geodesic line (one-parametric set of the quantum state vectors) that connects two state vectors $|\psi_0\rangle$
and $|\psi_1\rangle$ can be defined as their linear combination
\begin{eqnarray}\label{geodesic}
|\psi(\xi)\rangle=C\big[(1-\xi)|\psi_0\rangle+\xi|\psi_1\rangle
\re^{\ri \phi}\big],
\end{eqnarray}
where $\xi$ is a real parameter changing from $0$ to $1$.
This definition is similar to the definition of a direct line connecting two points ${\bf r}_0$ and ${\bf r}_1$ in Euclidean space ${\bf r}=(1-\xi){\bf r}_0+\xi{\bf r}_1$.
However, in contrast to the classical case, in quantum case
the states $|\psi_0\rangle$ and
$\re^{\ri \phi_0}|\psi_0\rangle$ describe the same quantum state,
similarly, $|\psi_1\rangle$ and $\re^{\ri \phi_1}|\psi_1\rangle$
describe the same quantum state.
Therefore, we require that
geodesic lines defined between the states $|\psi_0\rangle$,
$|\psi_1\rangle$ and between the states
$\re^{\ri \phi_0}|\psi_0\rangle$, $\re^{\ri \phi_1}|\psi_1\rangle$ coincide.
This requirement is satisfied if we choose
\begin{eqnarray} \label{phi}
\re^{\ri \phi}={\langle\psi_1|\psi_0\rangle\over
|\langle\psi_1|\psi_0\rangle|}\,.
\end{eqnarray}
The normalization condition $\langle\psi(\xi)|\psi(\xi)\rangle=1$
gives
\begin{eqnarray}
C={1\over\sqrt{1-2\xi(1-\xi)(1-|\langle\psi_1|\psi_0\rangle|)}}\,.
\end{eqnarray}

Now, let us show that (\ref{geodesic}) is really a geodesic line. For this purpose, we calculate its length
and show that it is a minimal possible length.
The geodesic line (\ref{geodesic}) is a one-parametric set of
states and there exist many possibilities to parameterize it.
One can
show that the length of the curve in quantum space does not depend
on the way of its parametrization.
To calculate the length
of the geodesic line it is convenient to write its equation as
follows:
\begin{eqnarray}\label{geodesic2}
|\psi(\theta)\rangle=C'\big[\sin({\theta/2})|\psi_0\rangle+\cos({\theta/2})|\psi_1\rangle
\re^{\ri \phi}\big],
\end{eqnarray}
here, a new parameter $\theta$ changes in the range $0 \leqslant\theta\leqslant \pi$ and the normalization constant reads
\begin{eqnarray}
C'={1\over \sqrt {1+
|\langle\psi_1|\psi_0\rangle|\sin\theta}}\,.
\end{eqnarray}
Let us stress once more that (\ref{geodesic}) and (\ref{geodesic2})
describe the same one-parametric family of quantum state vectors, namely, the geodesic line.
Comparing (\ref{geodesic}) and (\ref{geodesic2}) we find the relation between the parameters $\xi$ and $\theta$
\begin{eqnarray} \label{xitheta}
\xi={\tan(\theta/2)\over 1+\tan(\theta/2)}\,.
\end{eqnarray}
One can verify that substituting (\ref{xitheta}) into (\ref{geodesic}) we find (\ref{geodesic2}).

\begin{figure}[!t]
\begin{center}
\includegraphics[width=0.5\textwidth]{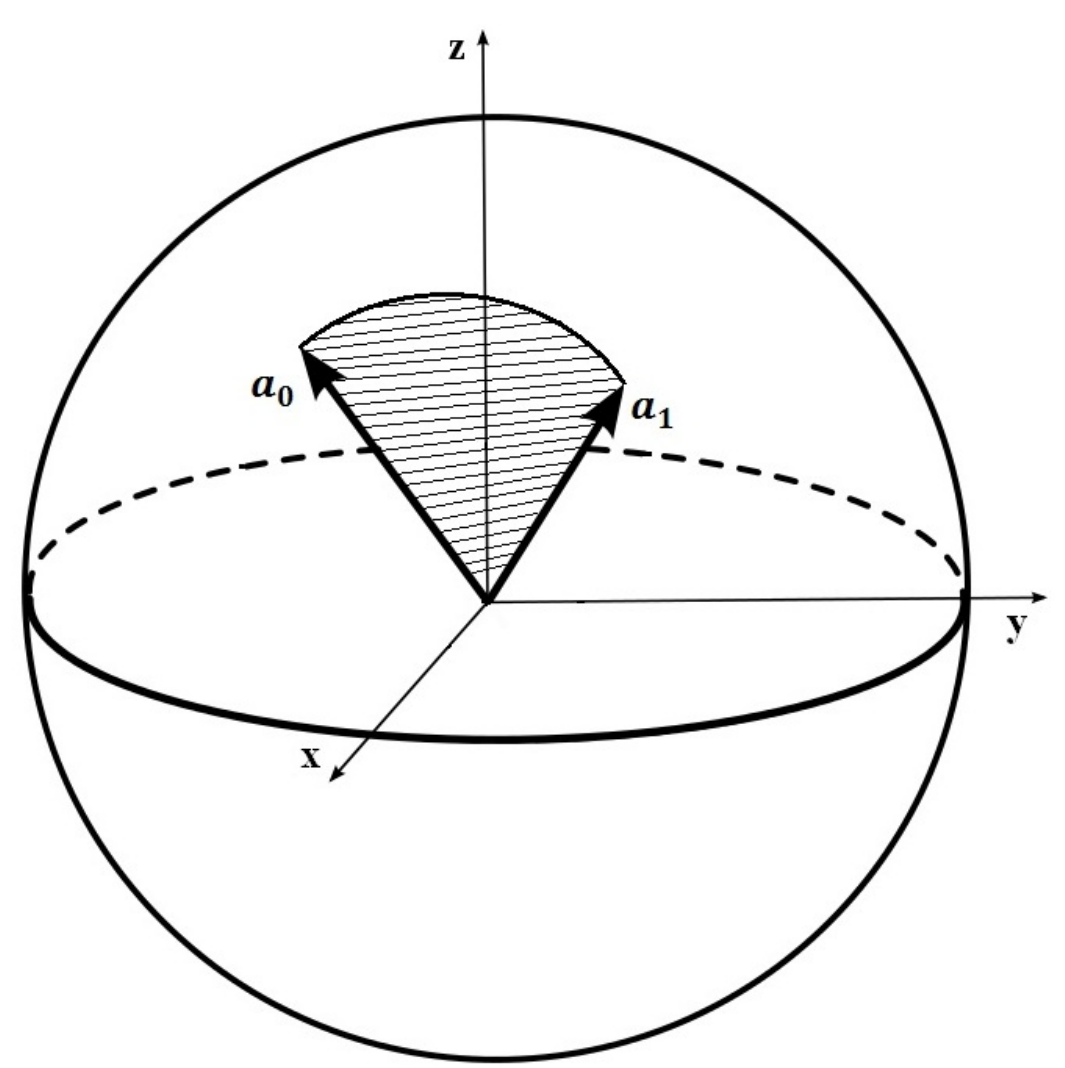}
\end{center}
\caption{Geodesic on Bloch sphere.  Vectors ${\bf a}_0$ and ${\bf a}_1$ on Bloch sphere correspond to $|\psi_0\rangle$ and
$|\psi_1\rangle$.}
\label{fig.1}
\end{figure}

Using (\ref{metric}) for the one-parameter set of states
(\ref{geodesic2}) we obtain
\begin{eqnarray}
\rd s={\gamma\over 2}{\sqrt{1-|\langle\psi_1|\psi_0\rangle|^2} \over
1+|\langle\psi_1|\psi_0\rangle|\sin\theta}\rd\theta.
\end{eqnarray}
Then, the length of the geodesic line connecting the states
$|\psi_0\rangle$ and $|\psi_1\rangle$ is
\begin{eqnarray}
s=\int \rd s=\int_0^{\pi}{\gamma\over 2}{\sqrt{1-|\langle\psi_1|\psi_0\rangle|^2} \over
1+|\langle\psi_1|\psi_0\rangle|\sin\theta}\rd\theta=\gamma\arccos|\langle\psi_1|\psi_0\rangle|.
\end{eqnarray}
Thus, this length is equal to the Wootters distance, that is, the length of geodesic
between two states (see figure~\ref{fig.1}). For $\gamma=2$, the Wootters distance  is equal to the angle between
vectors ${\bf a}_0$ and ${\bf a}_1$ on Bloch sphere which correspond to $|\psi_0\rangle$ and
$|\psi_1\rangle$, respectively. This angle is the minimal possible length of the curve on the Bloch sphere
connecting the states $|\psi_0\rangle$ and
$|\psi_1\rangle$.

In conclusion of this section, let us note that we can calculate
the length of the curve (\ref{geodesic}) connecting the states
$|\psi_0\rangle$ and $|\psi_1\rangle$ for an arbitrary phase
$\phi$. Then, the geodesic line can be defined as the one having a
minimal length. One can find that the minimal length is achieved for
$\phi$ given in (\ref{phi}) and is equal to the Wootters distance.

\section{Curvature}
The state vector of the quantum evolution belongs
to a one-parametric set of state vectors $|\psi(t)\rangle=\exp(-\ri Ht)|\psi_0\rangle$ generated
by the Hamiltonian of the system. The deviation of evolution state vector $|\psi(t)\rangle$
from the geodesic, connecting
the same two state vectors, is related with the curvature of quantum evolution.

In order to introduce the curvature as well as
the torsion, we consider the evolution in two stages. At first, we
consider the evolution during the time $\Delta t$ from an initial
state $|\psi_0\rangle$ to
\begin{eqnarray} \label{psip}
|\psi'\rangle=\re^{-\ri H\Delta t/\hbar}|\psi_0\rangle
\end{eqnarray}
and then the evolution during the time  $\Delta t'$ from $|\psi'\rangle$ to
\begin{eqnarray}\label{psi1}
|\psi_1\rangle=\re^{-\ri H\Delta t'/\hbar}|\psi'\rangle=\re^{-\ri H
(\Delta t+\Delta t')/\hbar}|\psi_0\rangle,
\end{eqnarray}
where $H$ is a time independent Hamiltonian. In this section,
without the loss of generality, we put \linebreak $\Delta t=\Delta t'$.

A deviation of the quantum evolution from the geodesic line
connecting $|\psi_0\rangle$ and $|\psi_1\rangle$ can be
characterized by the minimal distance between the state $|\psi'\rangle$
and the geodesic line $|\psi(\xi)\rangle$
\begin{eqnarray}
d^2=\min d^2(\xi)=
\min\gamma^2\left[1-|\langle\psi'|\psi(\xi)\rangle|^2\right].
\end{eqnarray}
The minimal value of this expression is achieved at $\xi=1/2$.
Taking into account the terms of order $(\Delta t)^4$ we find
\begin{eqnarray} \label{dmax}
d^2={\gamma^2\over4}\left[\langle(\Delta
H)^4\rangle-\langle(\Delta H)^2\rangle^2\right]{(\Delta
t)^4\over\hbar^4}={\gamma^2\over4}\kappa{(\Delta
t)^4\over\hbar^4}\,.
\end{eqnarray}
Here, the multiplier
\begin{eqnarray}
\kappa=\langle(\Delta H)^4\rangle-\langle(\Delta H)^2\rangle^2
\end{eqnarray}
can be called the curvature coefficient or curvature. It
is convenient to introduce a dimensionless curvature coefficient
\begin{eqnarray} \label{barcurv}
\bar \kappa={\langle(\Delta H)^4\rangle-\langle(\Delta
H)^2\rangle^2\over\langle(\Delta H)^2\rangle^2}\,.
\end{eqnarray}
For the first time, this result was obtained  within the framework of the study of the geometry
of quantum statistical interference in \cite{Brody96}.

Now, we show that the curvature of the quantum evolution can also
be obtained using the geometric treatment. For a small time, the
classical motion along a given curve can be treated as a motion
along the circle with radius $R$ for which we can write
\begin{eqnarray} \label{1R}
{1\over R}={2d\over(s/2)^2}\,,
\end{eqnarray}
where $s$ is the length of the curve between two neighboring
points on it, which can be considered as an arc of the circle, and
$d$ is the distance between the middle point of an arc and the
chord connecting these two points. Similarly to (\ref{1R}) we
define the radius of the curvature for the quantum evolution. In our
case, $d$ is given by (\ref{dmax}) and
\begin{eqnarray}
s=v2\Delta t=\gamma{\sqrt{\langle(\Delta H)^2\rangle}\over
\hbar}2\Delta t
\end{eqnarray}
is the length that a quantum system passes during the time
$2\Delta t$ of the evolution. Here, $v$ is the velocity of quantum
evolution given in (\ref{velocity}). As a result, we have
\begin{eqnarray}
{1\over R^2}={1\over\gamma^2}{\langle(\Delta
H)^4\rangle-\langle(\Delta H)^2\rangle^2\over\langle(\Delta
H)^2\rangle^2}={\bar \kappa\over\gamma^2}\,.
\end{eqnarray}

\section{Torsion}
Torsion is related with the deviation of the evolution state vector from the plane of evolution
(a two-dimensional subspace) at a given time.

In order to find torsion, we consider
the evolution in the two stages given by (\ref{psip}) and
(\ref{psi1}). Two vectors $|\psi_0\rangle$ and $|\psi'\rangle$ that
form the first stage
define the plane of evolution. Using these vectors we can
construct the orthogonal ones
\begin{eqnarray}
|\phi_1\rangle={1\over
\sqrt{2(1+a)}}\left(|\psi_0\rangle+\re^{-\ri \alpha}|\psi'\rangle\right),\\
|\phi_2\rangle={1\over
\sqrt{2(1+a)}}\left(|\psi_0\rangle-\re^{-\ri \alpha}|\psi'\rangle\right),
\end{eqnarray}
where $a$ and $\phi$ are defined by
$\langle\psi_0|\psi'\rangle=a\re^{\ri \alpha}$. Then, the unit operator
in a two-dimensional subspace spanned by $|\phi_1\rangle$ and
$|\phi_2\rangle$ is
\begin{eqnarray}
\hat
I_2=|\phi_1\rangle\langle\phi_1|+|\phi_2\rangle\langle\phi_2|.
\end{eqnarray}
Note that this is the projection operator of an arbitrary state
vector on a two-dimensional subspace.

In order to find the deviation of the state vector
$|\psi_1\rangle$ obtained at the second stage from the plane of evolution, we calculate the mean
value of $\hat I_2$
\begin{eqnarray}
I_2=\langle\psi_1|\hat I_2|\psi_1\rangle
=|\langle\phi_1|\psi_1\rangle|^2+|\langle\phi_2|\psi_1\rangle|^2.
\end{eqnarray}
When $I_2=1$, then $|\psi_1\rangle$ belongs to the subspace spanned
by $|\phi_1\rangle$ and $|\phi_2\rangle$. It means that three state vectors
$|\psi_0\rangle$, $|\psi'\rangle$, $|\psi_1\rangle$ belong to the
same plane (the two-dimensional subspace) of evolution and thus
the torsion is zero. The expression $1-I_2$ gives the magnitude of
the torsion. Considering small $\Delta t$ and $\Delta t'$ and
taking into account the terms up to the fourth order, we find
\begin{eqnarray}
1-I_2=\left[\langle(\Delta H)^4\rangle-\langle(\Delta
H)^2\rangle^2-{\langle(\Delta H)^3\rangle^2\over\langle(\Delta
H)^2\rangle}\right]{\Delta t^2(\Delta t+\Delta t')^2\over
4\hbar^4}\,.
\end{eqnarray}
The coefficient
\begin{eqnarray}
\tau=\langle(\Delta H)^4\rangle-\langle(\Delta
H)^2\rangle^2-{\langle(\Delta H)^3\rangle^2\over\langle(\Delta
H)^2\rangle}
\end{eqnarray}
does not depend on $\Delta t$ and $\Delta t'$ and can be called
the torsion coefficient. For simplicity, we put $\Delta t=\Delta
t'$ and then
\begin{eqnarray}
1-I_2=\tau{\Delta t^4\over \hbar^4}\,.
\end{eqnarray}
Similarly to the dimensionless curvature coefficient, we introduce
a dimensionless torsion coefficient
\begin{eqnarray}\label{bartor}
\bar\tau={\tau\over\langle(\Delta H)^2 \rangle^2} =
{\langle(\Delta H)^4\rangle-\langle(\Delta
H)^2\rangle^2\over\langle(\Delta H)^2\rangle^2}-{\langle(\Delta
H)^3\rangle^2\over\langle(\Delta H)^2\rangle^3}=\bar
\kappa-{\langle(\Delta H)^3\rangle^2\over\langle(\Delta
H)^2\rangle^3}\,.
\end{eqnarray}

Now, let us show that $1-I_2$ has a geometrical meaning, namely, it
is proportional to the squared distance of the state
$|\psi_1\rangle$ to the plane of quantum evolution spanned by
$|\phi_1\rangle$ and $|\phi_2\rangle$. The distance between a
given state $|\psi_1\rangle$ and the plane is equal to the distance between
$|\psi_1\rangle$ and the normalized projection of this vector onto
the plane. The normalized projection of $|\psi_1\rangle$ on the
plane is
\begin{eqnarray}
|\tilde\psi_1\rangle=c \hat I_2|\psi_1\rangle,
\end{eqnarray}
where from the condition
$\langle\tilde\psi_1|\tilde\psi_1\rangle=1$, we find
$c=(\langle\psi_1|\hat I_2\hat I_2|\psi_1\rangle)^{-1/2}=
(\langle\psi_1|\hat I_2|\psi_1\rangle)^{-1/2}$. Here, we use that
$(\hat I_2)^2=\hat I_2$. Then, the squared distance between the
state $|\psi_1\rangle$ and the plane is
\begin{eqnarray}
d^2=\gamma^2\left(1-|\langle\psi_1|\tilde\psi_1\rangle|^2\right)=
\gamma^2\left(1-|\langle\psi_1|\hat
I_2|\psi_1\rangle|\right)=\gamma^2(1-I_2),
\end{eqnarray}
where we use that $|I_2|=I_2$.

\section{Discussion}
In this paper, we have obtained the curvature
and torsion coefficients (\ref{barcurv}) and (\ref{bartor}) for
the quantum evolution which is governed by a time independent
Hamiltonian. In this case, the curvature and torsion coefficients
are constant.

The evolution is going along the geodesic when $\bar
\kappa=0$  or explicitly
\begin{eqnarray}\label{Eqk0}
\langle(\Delta H)^4\rangle-\langle(\Delta H)^2\rangle^2=0.
\end{eqnarray}
Introducing operator $\hat{A}= (\Delta H)^2$ we rewrite this
condition in the form $\langle(\Delta \hat A)^2\rangle=0$. Then, we
find that (\ref{Eqk0}) is equivalent to the equation $\Delta \hat
A|\psi\rangle=0$ which explicitly reads
\begin{eqnarray} \label{Eig}
(\Delta H)^2|\psi\rangle=\langle(\Delta H)^2\rangle|\psi\rangle.
\end{eqnarray}
The solution of this equation is
\begin{eqnarray} \label{psi0}
|\psi\rangle={1\over\sqrt
2}\left(|E_1\rangle+\re^{\ri \alpha}|E_2\rangle\right),
\end{eqnarray}
where $\alpha$ is an arbitrary phase, $|E_1\rangle$ and
$|E_1\rangle$ are two eigenstates of the Hamiltonian $H$ with
eigenenergies $E_1$ and $E_2$. Considering (\ref{psi0}) as an initial state for the time
dependent state that evolves along the geodesic we find
\begin{eqnarray} \label{psit}
|\psi(t)\rangle={1\over\sqrt
2}\left(\re^{-\ri E_1t/\hbar}|E_1\rangle+\re^{\ri \alpha}\re^{-\ri E_2t/\hbar}|E_2\rangle\right).
\end{eqnarray}
One can find that for an arbitrary time, the evolution state vector (\ref{psit}) satisfies the
equation~(\ref{Eig}) and
the curvature for this evolution is zero.
It is worth stressing that the state vector of the geodesic
evolution contains only two eigenstates of the Hamiltonian and
lies in a two-dimensional subspace.

Let us show that the torsion of the geodesic is zero. Using
(\ref{Eig}) we have
\begin{eqnarray}
\langle(\Delta H)^3\rangle=\langle\psi|\Delta H(\Delta
H)^2|\psi\rangle=\langle(\Delta H)^2\rangle\langle\psi|\Delta
H|\psi\rangle=0.
\end{eqnarray}
Then, according to (\ref{bartor}) and taking into account that for the
geodesic $\bar\kappa=0$ we find that the torsion $\bar\tau=0$.

Let us verify that for two-dimensional space the torsion given by
(\ref{bartor}) is zero because this is abiding by the definition. The
most general Hamiltonian for a two-dimensional case reads
\begin{eqnarray}\label{HtooD}
H= \omega(\mbox{\boldmath $\sigma$}{\bf n}) + \epsilon,
\end{eqnarray}
where $\bf n$ is a unit vector.
Note that the curvature and torsion depend on $\Delta H$, where $\epsilon$ is
canceled. So, without loss of generality we put $\epsilon=0$. Then,
using the properties of Pauli
matrices for Hamiltonian (\ref{HtooD}) with $\epsilon=0$  we find the following results
$\langle(\Delta
H)^2\rangle=\omega^2-\langle H\rangle^2$,
$\langle(\Delta
H)^4\rangle-\langle(\Delta H)^2\rangle^2
=4\langle H\rangle^2\langle(\Delta
H)^2\rangle$ and $\langle(\Delta
H)^3\rangle=-2\langle H\rangle\langle(\Delta
H)^2\rangle$. Then, one can find that the torsion (\ref{bartor}) in
the two-dimensional case is always zero.
For curvature in this case, we have $\bar\kappa=4\langle H\rangle^2/\langle(\Delta
H)^2\rangle$. Thus, in two-dimensional case the quantum evolution is going along
the geodesic line when
$\langle H\rangle=\omega\langle(\mbox{\boldmath $\sigma$}{\bf n})\rangle=0$.

In conclusion, let us note an interesting fact which follows from
(\ref{bartor}). Namely, for symmetric states when $\langle(\Delta
H )^3\rangle=0$, we find that $\bar\kappa=\bar\tau$. It means that
the curvature and torsion during the evolution of symmetric states
are strongly related.

Finally, we would like to note that the curvature and torsion, presented in this paper, are interesting on their own rights. They can be used for the study of evolution of different quantum systems.
In this paper, we presented a simple example of quantum system, namely spin in magnetic field.  For the system, we find curvature and torsion during evolution. Of course, this example can be considered as a simple demonstration. It is also interesting to study curvature and torsion for a many-spin system during the evolution, in particular for a two-spin system. We think that such characteristics of quantum evolution as curvature and torsion are useful for the study of brachistochrone problem and entanglement. Note also that in this paper we considered curvature and torsion for a time independent Hamiltonian. Of course, there appears a question regarding generalization of these characteristics on time dependent Hamiltonian. This question is worth to be studied separately.

\section*{Acknowledgement}
We would like to thank Dr. Yu. Krynytskyi for constructive
discussions and useful comments.
We also thank the Members of Editorial Board  for the possibility to present our results in a special issue of
Condensed Matter Physics dedicated to Prof. Holovatch's 60th birthday.
We admire his activities in various fields of science and in scientific life and wish him bright ideas in future.

\ukrainianpart

\title{Геометричні характеристики квантової еволюції: кривизна та кручення}
\author{Г.П. Лаба \refaddr{label1}, В.М. Ткачук \refaddr{label2}}
\addresses{
\addr{label1} Кафедра прикладної фізики і наноматеріалознавства, Національний університет ``Львівська політехніка'',\\ вул. Устияновича, 5, 79013 Львів, Україна
\addr{label2} Кафедра теоретичної фізики, Львівський національний університет імені Івана Франка, \\ вул. Драгоманова, 12, 79005 Львів, Україна
}

\makeukrtitle

\begin{abstract}
\tolerance=3000%
Ми вивчаємо характеристики квантової еволюції, такі як кривизна та кручення. Кривизна показує відхилення вектора стану під час квантової еволюції від геодезичної лінії. Кручення визначає відхилення вектора стану від площини еволюції (двовимірний підпростір) в заданий момент часу.

\keywords кривизна, кручення, квантова еволюція, геометрія простору квантових станів

\end{abstract}

\end{document}